\documentclass[10pt,twocolumn,superscriptaddress,floatfix,showpacs,nobalancelastpage,longbibliography]{revtex4-1}

\usepackage{array,longtable}
\newcolumntype{L}{>{\tiny $}p{0.33\columnwidth}<{$}}
\newcolumntype{M}{>{\scriptsize $}p{0.33\columnwidth}<{$}}
\newcolumntype{N}{>{\scriptsize $}p{0.43\columnwidth}<{$}}
\usepackage{amsmath}
\usepackage{amssymb}
\usepackage{amsfonts}
\usepackage{graphicx}
\usepackage{hyperref}

\usepackage{float}
\usepackage{graphics}

\allowdisplaybreaks[1]

\newif\ifhyper
\hypertrue
\ifhyper
\hypersetup{
   citecolor = {green},
   colorlinks = {true}, 
   urlcolor = {blue} 
}
\fi

\begin{document}

\title{Investigation of the chiral antiferromagnetic Heisenberg model using PEPS}

\author{Didier Poilblanc}
\affiliation{Laboratoire de Physique Th\'eorique, C.N.R.S. and Universit\'e de Toulouse, 31062 Toulouse, France}

\date{\today}

\begin{abstract}

 A simple spin-$1/2$ frustrated antiferromagnetic Heisenberg model (AFHM) on the square lattice 
 -- including chiral plaquette cyclic terms --
 was argued [Anne E.B.~Nielsen, Germ\'an~Sierra and J.~Ignacio~Cirac, Nature Communications {\bf 4}, 2864 (2013)] to host a bosonic Kalmeyer-Laughlin (KL) fractional quantum Hall ground state [V. Kalmeyer and R. B. Laughlin,
Phys. Rev. Lett. {\bf 59}, 2095 (1987)]. Here, 
 we construct generic families
 of chiral projected entangled pair states (chiral PEPS) with low bond dimension ($D=3,4,5$)  
 which, upon optimization, provide better
 variational energies than the KL ansatz. The optimal $D=3$ PEPS exhibits 
 chiral edge modes described by the Wess-Zumino-Witten
 $SU(2)_1$ model, as expected for the KL spin liquid. However, we find evidence that, 
 in contrast to the KL state, the PEPS spin liquids have
 power-law dimer-dimer correlations and exhibit a gossamer long-range tail in the spin-spin correlations. 
 We conjecture that these features are genuine to {\it local} chiral AFHM on bipartite lattices.

 \end{abstract}
\maketitle


{\it Introduction.} 
Topological order (TO) has been rationalized in the last few decades~\cite{Wen1990,Wen2013} as a new type of order in two dimensions (2D), 
beyond the well-known 
Ginzburg-Landau paradigm. Importantly, it is at the heart of the rapidly expanding field of quantum computing~\cite{Kitaev2003}.  
The fractional quantum Hall (FQH) state of the 2D electron gas~\cite{Stormer1983} is the first topological ordered state discovered. The simple Laughlin wave function provides a beautiful qualitative understanding of the physics of the Abelian FQH state at filling fraction $\nu=1/m$ as an incompressible fluid~\cite{Laughlin1983},
while more involved wave functions can also describe non-Abelian FQH states~\cite{Moore1991,Read1999,Repellin2015}. 
It revealed 
the emergence of fractional excitations, the anyons, a key feature of TO~\cite{Wen1990}. Anyons carry fractional charge~\cite{Laughlin1983} 
as well as Abelian~\cite{Halperin1984} or non-Abelian statistics~\cite{Wen1991b,Moore1991}.
An important feature of FQH states is the existence of a bulk gap and chiral modes providing
unidirectional transport on the edge~\cite{Wen1991a,Wen1992}.
More precisely, their edge physics
can be described by chiral $SU(2)_k$ Wess-Zumino-Witten
(WZW) Conformal Field Theory (CFT)~\cite{WZWreview1988}. 
Recently, a matrix product state (MPS) representation of the FQH states~\cite{Estienne2013a,Estienne2015} enabled to
probe their physical properties with unprecedented numerical accuracy.

In a pioneering work~\cite{Kalmeyer1987}, Kalmeyer and Laughlin (KL) have extended the notion of FQH state to the lattice.
When localized on the lattice, the bosonic $\nu=1/2$ Laughlin state gives rise to a spin-$1/2$ chiral spin liquid
(CSL)~\cite{WWZ1989},
closely related to the resonating valence bond (RVB) state of high-Tc superconductivity~\cite{Anderson1973}.
Recently, fractional Chern insulators~\cite{Levin2009,Repellin2014,Maciejko2015} have set up a new route to realize FQH physics on the lattice. 

\begin{figure}[htbp]
\begin{center}
\includegraphics[width=\columnwidth,angle=0]{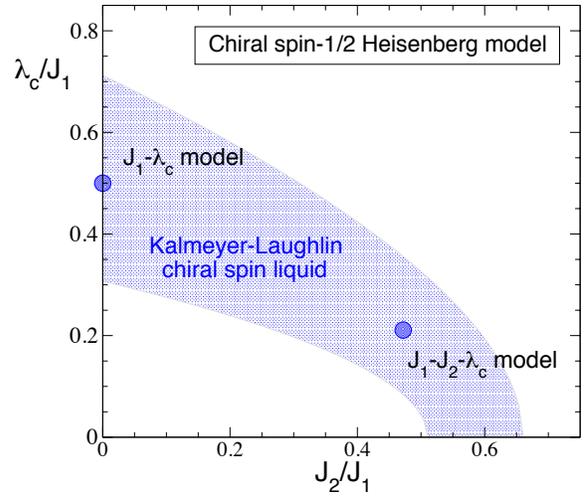}
\caption{[Color online] Schematic phase diagram of the chiral AFHM drawn from Ref.~\protect\cite{Nielsen2013} as a function of magnetic frustration $J_2/J_1$ and (relative) amplitude of the chiral interaction $\lambda_c/J_1$.
The KL nature and the boundary of the chiral spin liquid phase (blue region) was only accessed from small cluster
calculations. The location in parameter space of the two models studied here are shown by (blue) dots.}
\label{FIG:PhaseDiag}
\end{center}
\end{figure}

Whether simple local lattice Hamiltonians can host chiral spin liquid ground states~\cite{WWZ1989} is one of the key issues
that determine whether or not such topological phases could be realized experimentally.  
The original innovative proposal by KL that the GS of the frustrated triangular spin-$1/2$ antiferromagnetic Heisenberg model (AFHM)
is a CSL turned out not to be correct, the GS of this model being
magnetically ordered. However, Bauer et al.~\cite{Bauer2014} showed recently that, on the kagome lattice (2D lattice of corner-sharing triangles), the GS of the Hamiltonian $H=\sum_{\triangle(ijk)} {\bf S}_i \cdot ({\bf S_j}\times {\bf S_k})$, sum of the chiral spin interaction 
over all triangles $\triangle(ijk)$, has the universal properties of the $\nu=1/2$  Laughlin state.
This CSL was shown to be exceptionally robust under the addition of an extra nearest-neighbor Heisenberg-like interaction (defining a generic ``chiral AFHM"), even of large magnitude. 

\begin{table}[htbp]
\caption{Numbers of independent $SU(2)$-symmetric tensors for the four different 
virtual spaces we consider, $D\le 5$. The third (fourth) column gives the number of $A_1$ ($A_2$) tensors and the last column the total number of tensors in the $A$ ansatz. Note that all four types of ans\"atze exhibit a gauge-$\mathbb Z_2$ symmetry associated to the conserved parity of the number of spin-$\frac{1}{2}$ on the $z=4$ bonds. }
\begin{center}
 \begin{tabular}{@{} ccccc@{}}
   \hline 
   \hline
    ${\cal V}$ &  $D$ &$A_R^{(A_1)}$ &   $A_I^{(A_2)}$ & Total \#  \\ 
    \hline
    $\frac{1}{2}\oplus 0$&3& 2&  1& 3 \\
    $\frac{1}{2}\oplus 0\oplus 0$&4& 8&  4& 12 \\
     $\frac{1}{2}\oplus \frac{1}{2}\oplus 0$&5& 10&  8& 18 \\
    $\frac{1}{2}\oplus 0\oplus 0\oplus 0$&5& 21&  12& 33 \\
     \hline 
        \hline 
      \end{tabular}
\end{center}
\label{TABLE:numbers}
\end{table}%

Another alternative approach has been pursued, trying to construct ``parent Hamiltonians" for the Abelian~\cite{Schroeter2007,Thomale2009}
and non-Abelian~\cite{Greiter2014,Glasser2015} CSL.
Using 
a re-writing of the wave function as a correlator of a $1+1$ chiral CFT~\cite{Nielsen2011,Nielsen2012},
the simplest \hbox{spin-$\frac{1}{2}$} parent Hamiltonian on the square lattice obtained by Nielsen et al.~\cite{Nielsen2013} consists of interactions between all pairs and triples of spins in the system.
Since long-range interactions might be hard to achieve experimentally in e.g. cold atom systems~\cite{Nielsen2014}, the authors argue that 
a similar (Abelian) CSL phase is also hosted in a simplified {\it local} Hamiltonian where all the long-range parts of the interaction
have been set to zero~\cite{Nielsen2013}. We shall adopt here their local chiral AFHM which, introducing a slightly different parametrization, reads:
\begin{eqnarray}
H&=&J_1\sum_{\big< i,j\big>} {\bf S}_i\cdot{\bf S}_j
+J_2\sum_{\big<\big<k,l\big>\big>} {\bf S}_k\cdot{\bf S}_l\nonumber \\
&+&  \lambda_c \sum_{\square(ijkl)}i(P_{ijkl}-P_{ijkl}^{\,\,\,\,\,\,\,-1}) \, ,
\label{EQ:model}
\end{eqnarray} 
where the first (second) sum is taken over nearest-neighbor (next-nearest-neighbor) bonds and 
the last sum over all plaquettes of the square lattice. $P_{ijkl}$ makes a cyclic permutation 
of the four spins of every plaquette in e.g. the clockwise direction. $H$ breaks time reversal symmetry but preserves the global spin
$SU(2)$ symmetry. 
It is the analog for the square lattice of the chiral AFHM on the kagome lattice studied by Bauer et al.~\cite{Bauer2014}~:
The chiral interaction ${\bf S}_i \cdot ({\bf S_j}\times {\bf S_k})$ on the triangle is replaced here by its generalization on the plaquette
and magnetic frustration is introduced via competing $J_1$ and $J_2$ antiferromagnetic couplings.
A schematic phase diagram showing the (approximate) extension of the KL chiral spin liquid 
is provided for convenience in Fig.~\ref{FIG:PhaseDiag}. We shall here
focus on the two special points studied by Nielsen et al.~\cite{Nielsen2013} and located in Fig.~\ref{FIG:PhaseDiag}, supposedly in the CSL phase; $J_1=2$, $J_2=0$, $\lambda_c=1$ 
and $J_1=2\cos{(0.06\pi)}\cos{(0.14\pi)}\simeq 1.78$, $J_2=2\cos{(0.06\pi)}\sin{(0.14\pi)}\simeq 0.84$, 
$\lambda_c=2\sin{(0.06\pi)}\simeq 0.375$. Hereafter, we refer to the latter as the ``$J_1-\lambda_c$ model''  and the ``$J_1-J_2-\lambda_c$ model'', respectively. 

\begin{figure}[htbp]
\begin{center}
\includegraphics[width=\columnwidth,angle=0]{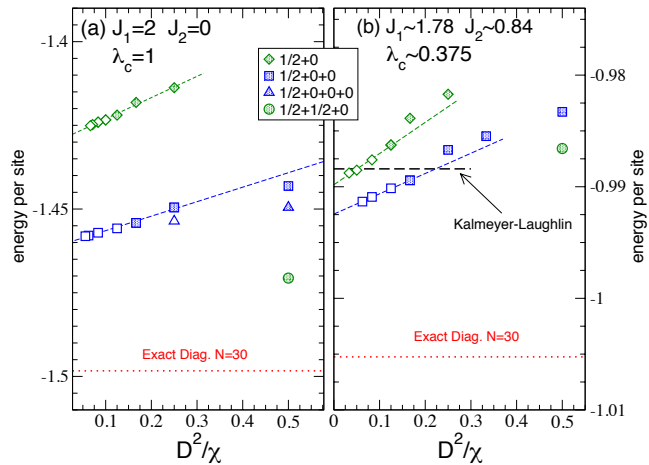}
\caption{
[Color online] Scaling of the iPEPS variational energies (per site) vs $D^2/\chi$
for the two local chiral Hamiltonians studied here;
(a) $J_1-\lambda_c$ model; (b) $J_1-J_2-\lambda_c$ model. 
The filled (open) symbols correspond to fully optimized (fixed) tensors as 
explained in the text.
A comparison 
with the exact energy (per site) of a 
$5\times 6$ torus~\cite{Nielsen2013} is shown. In (b) the variational energy
of the Kalmeyer-Laughlin (KL) spin liquid obtained by Monte Carlo~\cite{Nielsen2013} is also shown.
 }
\label{FIG:Ener}
\end{center}
\end{figure}

Our strategy to explore the physics of the above model is to use the 
tensor network framework~\cite{Cirac2009b,Cirac2012a,Orus2013,Schuch2013b,Orus2014}.
One of the motivation is to test whether some fundamental obstruction is at play that prevents to describe 
a gapped CSL phase with 2D tensor networks~\cite{Dubail2015}. Previous attempts using projected entangled pair states (PEPS) led 
to the discovery of {\it critical} CSL exhibiting chiral edge modes~\cite{Shuo2015,Poilblanc2015,Poilblanc2016}.
PEPS are ans\"atze
that approximate GS wave functions in terms of a unique site tensor $A_{\alpha\beta\gamma\delta}^{s}$,
where the greek indices label the states of the $D$-dimensional {\it virtual} spaces $\cal V$ attached to each site in the 
$z=4$ directions of the lattice,
and $s=\pm\frac{1}{2}$ is the $S_z$ component of the physical spin. The site tensors are then 
entangled together (i.e. contracted w.r.t. their virtual indices) to form a 2D tensor network. A priori, all the 
$2D^4$ coefficients of the site tensor can serve as parameters to optimize the variational GS energy.
However, the CSL bears a number of symmetry properties that greatly constrains the PEPS ansatz.
Recently, a classification of fully $SU(2)$-symmetric (singlet) PEPS was proposed~\cite{Mambrini2016} 
in terms of the irreducible representations (IRREP)
of the lattice point group ($C_{4v}$ in the case of the 2D square lattice). Since the CSL should be 
invariant under the combination of a reflection $\cal R$ w.r.t. to any crystalline direction ($x$, $y$, $x\pm y$) and time reversal
symmetry (i.e. complex conjugation), the simplest adequate PEPS site tensors 
have the form $A=A_R^{(A_1)} + i A_I^{(A_2)}$,
where the two real tensors $A_R^{(A_1)}$ and $A_I^{(A_2)}$ transform according to 
the $A_1$ (symmetric w.r.t. $\cal R$) and $A_2$ 
(antisymmetric w.r.t. $\cal R$) IRREP~\cite{Poilblanc2015,Poilblanc2016}.
These tensors have been tabulated in Ref.~\onlinecite{Mambrini2016} for $D\le 6$,
and their numbers for all virtual spaces $\cal V$ considered
in this work are listed in Table~\ref{TABLE:numbers}.
Following a previous study of the non-chiral frustrated AFHM~\cite{Poilblanc2017}, we consider a general superposition of all tensors of each class, the weights in the sum being considered as
variational parameters. As in the non-chiral case, the energy or observables can be computed directly in the thermodynamic limit using infinite-PEPS (iPEPS) 
corner transfer matrix (CTM) renormalization group (RG) techniques~\cite{Nishino1996,Nishino2001,Orus2009,Orus2012}, making advantage of simplifications 
introduced by the use of point-group symmetric tensors~\cite{Poilblanc2017}. 
At each RG step a truncation of the (hermitian) CTM is done 
keeping (at most) $\chi$ eigenvalues and preserving exactly the $SU(2)$ multiplet structure. 
Energy optimization~\cite{Corboz2016,Vanderstraeten2016,Liu2017} 
is performed using a conjugate gradient (CG) 
method~\cite{Numerical2007} up to a maximum 
$\chi=\chi_{\rm opt}$ and then, eventually, one takes the limit $\chi\rightarrow\infty$ (using a ``rigid" ansatz) by extrapolating the data~\cite{Poilblanc2017}.

We now turn to the results. In Fig.~\ref{FIG:Ener} we show the scaling of the iPEPS  energies vs $D^2/\chi$
for the two local chiral Hamiltonians studied here, and different choices of the virtual space
$\cal V$ up to $D\le 5$. Using linear fits, one obtains accurate variational energies 
in the $\chi\rightarrow\infty$ limit, apart from $D=5$ for which the
CTM RG converges to unphysical (pairs of) solutions beyond $\chi=2D^2$. 
The exact GS energies obtained on a small periodic 
30-site cluster~\cite{Nielsen2013} (expected to give a lower bound of the true thermodynamic values) provide a first reference,
showing that the iPEPS energies are remarkably accurate. For the second model in Fig.~\ref{FIG:Ener}(b),
we have compared our results to the variational energy of the KL ansatz computed with Monte Carlo~\cite{Nielsen2013}.
We find that, even for the smallest bond dimensions $D=3$ (${\cal V}=\frac{1}{2}\oplus 0$) and $D=4$ (${\cal V}=\frac{1}{2}\oplus 0\oplus 0$), 
the iPEPS energy is lower than the energy of the KL CSL. 
This provides solid arguments that these chiral $SU(2)$-invariant PEPS are very good variational states.
Hereafter we investigate further their edge and bulk properties and point out similarities and differences with the KL wave function.
\begin{figure}[htbp]
\begin{center}
\includegraphics[width=\columnwidth,angle=0]{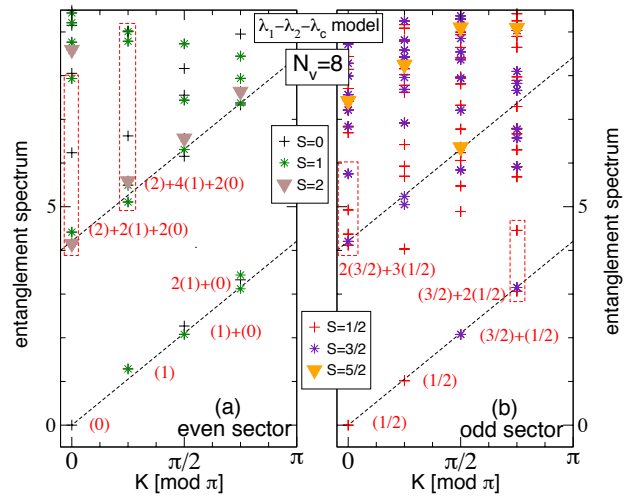}
\caption{[Color online]  Chiral entanglement spectra of the $D=3$ PEPS optimized for the $J_1-J_2-\lambda_c$ model (subtracting the GS energy for convenience) for $N_v=8$. The 
edge momentum $K$ is defined mod-$\pi$ since the $SU(2)$ generators are invariant under
only sublattice translations. 
Even (a) and odd (b) $\mathbb Z_2$ sectors are shown. The correct $SU(2)_1$
counting obtained for each quasi-degenerate group of levels at low energy (outlined by boxes when necessary) is indicated in red.}
\label{FIG:ES}
\end{center}
\end{figure}

{\it Chiral edge modes.} First, we have computed the entanglement spectrum (ES) of the optimized $D=3$ PEPS on an infinitely-long cylinder $\cal C$,
bi-partitioned into two semi-infinite half-cylinders ${\cal C}_{\rm L}$ and ${\cal C}_{\rm R}$,
${\cal C}={\cal C}_{\rm L}\cup {\cal C}_{\rm R}$.
This can be done exactly~\cite{Cirac2011} on cylinders with up to $N_v=8$ sites of circumference. Li and Haldane~\cite{Li2008} have conjectured 
that, in chiral topological states, there is a deep one-to-one correspondence between the true physical edge spectrum and the ES~\cite{Dubail2012a,Dubail2012b}. 
The ES is obtained
from the leading eigenvector of the finite-dimensional $D^{2N_v}\times D^{2N_v}$ transfer matrix of the cylinder, as 
originally proposed 
in Ref.~\onlinecite{Cirac2011}, and already applied to chiral spin liquids~\cite{Poilblanc2015,Poilblanc2016}.
The ES shown in Fig.~\ref{FIG:ES} as a function of the momentum $K$ along the cut clearly reveal the existence of well-defined chiral branches linearly dispersing as $E_K\sim v K$. One also sees quasi-degenerate groups of levels whose counting (in terms of $SU(2)$ multiplets) 
matches exactly the one of the $SU(2)_1$ WZW CFT~\cite{WZWreview1988}, as expected in a KL CSL phase~\cite{Herwerth2015}. Note that the ES of the optimized
PEPS is remarkably similar to the one obtained
for another studied chiral PEPS~\cite{Poilblanc2015,Poilblanc2016}, certainly belonging to the same $D=3$ chiral PEPS family, but far away in parameter space.   
Although, the same exact calculation cannot be realized for $N_v=8$ beyond $D=3$, we conjecture that the
$SU(2)_1$ chiral edge modes are genuine features of our chiral PEPS optimized for Hamiltonian (\ref{EQ:model}).

\begin{figure}[htbp]
\begin{center}
\includegraphics[width=\columnwidth,angle=0]{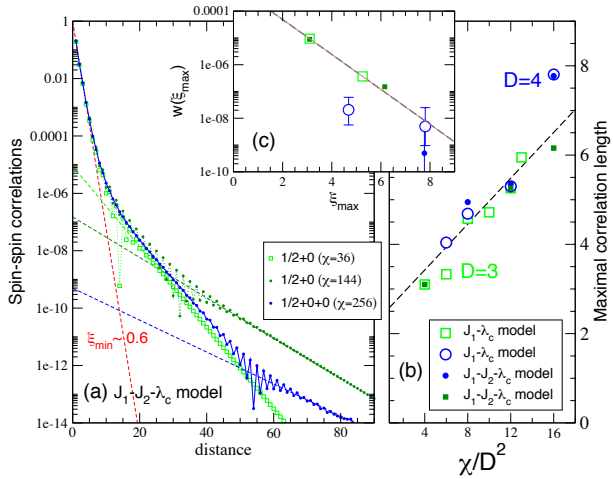}
\caption{[Color online]  (a) Absolute value of the spin-spin correlations vs distance (along some crystal axis direction $x$ or $y$)
for the $D=3$ and $D=4$ (optimized) chiral PEPS and different environment dimension $\chi$ 
(as shown in legends) on a semi-log plot. The dashed lines are fits according to exponential behaviors
of the short and long distance correlations.
(b) Largest correlation length $\xi_{\rm max}$ (obtained from the linear fits in (a)) vs $\chi/D^2$, for both model parameter sets. 
(c) ${\rm w} (\xi_{\rm max})$ versus $\xi_{\rm max}$ using the same symbols as in (b).}
\label{FIG:SS}
\end{center}
\end{figure}

{\it Bulk properties.} The KL CSL is expected to have short-range (spin-spin) correlations~\cite{Kalmeyer1987} as the bosonic $\nu=1/2$ FQH state it derives from. We investigate now 
the correlation functions of the PEPS ans\"atze, and establish important differences. We use the same definitions and 
CTM RG procedure as described in the study of the frustrated AFHM and focus on the two cases $D=3$ (${\cal V}=\frac{1}{2}\oplus 0$) and $D=4$ (${\cal V}=\frac{1}{2}\oplus 0\oplus 0$). 
Fig.~\ref{FIG:SS}(a) shows the spin-spin correlations vs distance on a semi-log plot.
At short distance, we observe a rapid exponential fall-off characteristic of the KL CSL. However our data clearly show additional exponential tails 
with much larger characteristic length but with much smaller weight. In other words, we can parametrize the  correlation function vs distance as 
\begin{equation}
C_S(d)=\sum_{\xi_{\rm min}\le\xi\le\xi_{\rm max}} {\rm w}(\xi)\exp{(-d/\xi)}\, ,
\end{equation}
where the short distance decay is characterized by ${\rm w}(\xi_{\rm min})\simeq 1$ while, at long distance, 
the slower decay $\exp{(-d/\xi_{\rm max})}$
takes over with $\xi_{\rm max}\gg \xi_{\rm min}$ and ${\rm w}(\xi_{\rm max})\ll 1$. In fact, we think that $\xi_{\rm max}\rightarrow\infty$
when $\chi\rightarrow \infty$ (see Fig.~\ref{FIG:SS}(b)) while, simultaneously, ${\rm w}(\xi_{\rm max})$ goes very rapidly to zero. 
If, as suggested in Fig.~\ref{FIG:SS}(c), ${\rm w}(\xi)\propto \exp(-\xi/\lambda)$, where
$\lambda\simeq 0.7\sim \xi_{\rm min}$, $C_S(d)$ will show a typical stretched exponential form at long distance,
$C_S(d)\sim (d/\lambda)^\frac{1}{4}\exp{\{-(d/\lambda)^\frac{1}{2}\}}$.
In any case, $C_S(d)$ should exhibit a ``gossamer tail'' which decays slower than any single exponential function. 

\begin{figure}[htbp]
\begin{center}
\includegraphics[width=\columnwidth,angle=0]{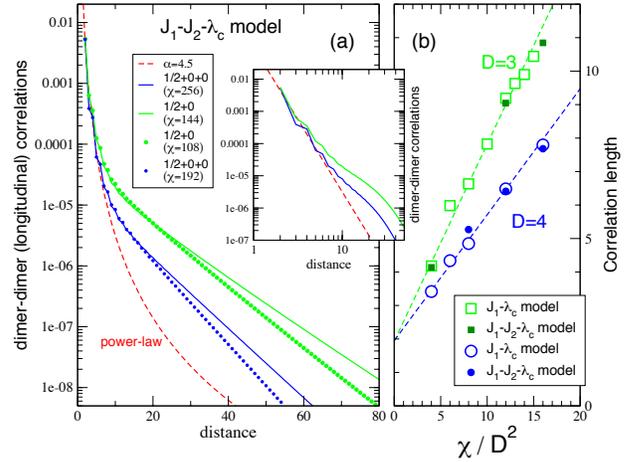}
\caption{[Color online]  (a) Absolute value of the dimer-dimer correlations vs distance $d$ (along some crystal axis direction)
for the $D=3$ and $D=4$ (optimized) chiral PEPS and different environment dimension $\chi$ 
(as shown in legends) on semi-log and log-log (inset) plots. The dashed (red) curve is a power-law $d^{-\alpha}$ fit.
(b) Correlation lengths obtained from the fits of the long distance correlations, shown vs $\chi/D^2$, for both model parameter sets. }
\label{FIG:DD}
\end{center}
\end{figure}

The dimer-dimer correlations are shown in Fig.~\ref{FIG:DD}(a). The asymptotic long-distance behaviors can always 
be fitted as exponential decays.
The correlation lengths extracted from the fits are found to diverge linearly with $\chi$, for both model studied, as shown 
in Fig.~\ref{FIG:DD}(b). At short distance, the data are better fitted as a power law $d^{-\alpha}$, although 
with a large exponent $\alpha\simeq 4.5$,
rather than as an exponential. Thus, the power-law behavior takes over at all distances when $\chi\rightarrow\infty$. This suggests a 
form of emerging $U(1)$-gauge symmetry typical of dimer liquids~\cite{Rokhsar1988} or RVB states~\cite{Albuquerque2010,Tang2011,Poilblanc2012,Schuch2012} on bipartite lattices. 

{\it Summary and outlook.} Using a previous symmetry classification of $SU(2)$-invariant PEPS we have constructed simple families of
chiral PEPS on the square lattice.
Using iPEPS supplement by a CG algorithm, we have optimized these PEPS w.r.t the local chiral (frustrated) AFHM, believed to host
a CSL phase of the same class as the $\nu=1/2$ bosonic FQH liquid. The energy optimizations reveal very competitive ans\"atze
(better than the KL ansatz) even for small bond dimensions $D=3$ or $D=4$. As expected in such a CSL phase, we find clear evidence of 
$SU(2)_1$ chiral edge modes. However, bulk properties turned out to have fundamental differences compared to a FQH gapped liquid~:
although spin-spin and dimer-dimer correlations seem qualitatively different, both seem to reveal long-range behaviors. 
Although, detailed data have been provided for two particular points in parameter space, a similar behavior 
has also been found between those two points. 
We conjecture that this may well be realistic features of the GS of (\ref{EQ:model}) which would host in fact a {\it critical} CSL.
Certainly, this does not contradict the results of Ref.~\onlinecite{Nielsen2013} showing that, on small clusters, the KL state is an extremely 
good ansatz for (\ref{EQ:model}).
Indeed, the short-range properties of our critical chiral PEPS are also likely to be extremely close to those of the KL state so that
only the long-distance properties can distinguish them. Interestingly, it was proved
that any strictly short-range quadratic parent Hamiltonian for chiral {\it free} fermions is gapless~\cite{Dubail2015}. It may well be that this extends to interacting local Hamiltonians, in agreement with our findings. 
This would also agree with the fact that the CFT wave function derived
using the null vectors of $SU(2)_1$~\cite{Nielsen2012,Nielsen2013},
i.e. the KL state, has a parent $H$ that is long range. 

\begin{acknowledgements}
This project is supported by the TNSTRONG
ANR grant (French Research Council).  This work was granted access to the HPC resources of CALMIP supercomputing center under the allocation 2017-P1231. I acknowledge inspiring conversations with Fabien Alet, Sylvain Capponi, Ignacio Cirac, Matthieu Mambrini, Anne Nielsen, Pierre Pujol, German Sierra and Norbert Schuch. I am also grateful to Anne Nielsen for providing the variational energy of the KL state to compare to. 

\end{acknowledgements}

\bibliography{bibliography}



\end{document}